\documentstyle[prd,aps,floats,times]{revtex}
\draft
\input epsf
\def\beqn{\begin{equation}}
\def\eeqn{\end{equation}}
\def\beq{\begin{equation}}
\def\eeq{\end{equation}}
\def\beqa{\begin{eqnarray}}
\def\eeqa{\end{eqnarray}}
\def\etal{et. al.~}

\begin{document}
\title{Quintessential Inflation with Dissipative Fluid}
\author{A. A. Sen$^{\dagger}$,~Indrajit Chakrabarty$^{\ddagger}$ and 
T. R. Seshadri$^{\star}$}
\address{Harsish-Chandra Research Institute, Chhatnag Road, Jhusi. 
Allahabad 211 019 
India}
\maketitle

\begin{abstract}
We have investigated cosmological models with a self-interacting scalar 
field and a dissipative matter fluid as the sources of 
matter. Different variables are expressed in terms of a {\it generating 
function}. Exact solutions are obtained 
for one particular choice of the {\it generating function} 
The potential corresponding to this generating function is a standard
tree-level potential arising in the perturbative regime in 
quantum field theory.
With suitable choice of parameters, the scale factor 
in our model exhibits both inflationary behaviour in the early universe
as well as an 
accelerating phase at late times with a decelerating period in 
between. It also satisfies the constraints for primeval nucleosynthesis and 
structure formation and seems to solve the cosmic coincidence problem. 
The solution exhibits a attractor nature towards a asymptotic de-sitter 
universe.
 
\end{abstract}

\pacs{PACS Number(s): 04.20Jb, 98.80Hw \hfill MRI-PHY/P20000531}
\vskip 2pc

\section{Introduction}

A number of recent observations \cite{flat} suggest that the 
$\Omega_m$, the ratio of the matter density(baryonic+dark) to 
the critical density, is significantly less than unity suggesting 
that either the universe is open or that there is some other 
sources of this missing energy which makes $\Omega_{total}\sim 1$. 
The recent findings of BOOMERANG experiments 
\cite{BOOM} strongly suggests the second possibility of a 
flat universe. At the same time, the measurements of the 
luminosity-redshift relations observed for the 50 newly discovered 
type Ia supernova with redshift $z>0.35$ \cite{super}  indicate 
that at present the universe is expanding in an accelerated fashion 
suggesting a net negative pressure for the universe.

Initial suggestions   
were to identify this missing energy density  to a 
cosmological constant $\Lambda$\cite{Rbah,Rsah}.
For a flat matter dominated universe with $\Lambda$ in Einstein gravity 
observations strongly suggest $\Omega_\Lambda\sim 0.72$. 
However, this possibility that 
$\Lambda$ could be the dominant energy density has the drawback  
that the energy scale involved is lower 
than normal energy scale predicted by the most particle physics models
by a factor of $\sim 10^{-123}$. 
An alternative source of energy density that may be admissible for 
this acceleration could be a dynamical 
$\Lambda$\cite{Rcal} in the form of a scalar field with some self 
interacting 
potential \cite{Rpeebetc}. If the energy density of this kind of source
varies slowly with time, it mimics 
an effective cosmological constant. The idea of this  candidate,
called {\it quintessence} \cite{Rcal}, is borrowed from the 
inflationary paradigm 
of the early universe. The difference, however, is that this new field 
evolves at a much 
lower energy scale. The energy density of this field, though dominant 
at present epoch, must remain sub-dominant at very early stages 
and should have
evolved in such a way that it becomes comparable to the matter density
$\rho_m$ now. When quintessence is modeled using a minimally 
coupled scalar field, in general, parameters need to be fine-tuned
so as to ensure that $\rho_m$ and $\rho_{\phi}$ are of the same order today.
This fine tuning problem has been termed as the cosmic coincidence problem. 
A new form 
of quintessence field called ``{\it tracker field}''\cite{Rzla} 
has been 
proposed to solve the cosmic coincidence problem. It has an equation
of motion with an attractor like solution in the sense that for a 
wide range of initial conditions the equation of motion converges 
to the same solution. 

There are a number of quintessence models which have been suggested 
and most of these involve scalar fields with minimal coupling
with potentials dominating over the kinetic energy of the field.     
A purely exponential potential is one of the widely
studied cases \cite{Rfer}. Inspite of the several advantages 
the energy density is not enough to make up for the missing part.
Inverse power law is another form of the 
potential (\cite{Rpeebetc}-\cite{Rzla}) 
that has
been considered extensively for quintessence models, in particular, for
solving the cosmic coincidence problem. Though many of the problems
are resolved successfully with this potential, the predicted 
value for the equation of state for the quintessence field, $\gamma_Q$,
is not in good agreement with the 
observed results. In search of suitable models that would 
eliminate the problems, new types of potentials, like $V_0[\cos h
\lambda\phi-1]^p$\cite{Rsah2} and $V_0\sin h (\alpha\sqrt
k_0\Delta\phi)^\beta$\cite{Rsah,Rlop} have been considered, which have
asymptotic forms like the inverse power law or exponential ones.
Different physical considerations have lead to the study of 
other types of the potentials also\cite{Ruzan}. 
Recently Saini {\it et al}  \cite{Rsai} have reconstructed the 
potential 
in context of general relativity and minimally coupled quintessence 
field from the expression of the luminosity distance $d_L(z)$
as function of redshift obtained from the observational data.
However, none of these potentials are entirely free of
problems. Hence, there is still
a need to identify appropriate potentials to explain
current observations \cite{Rfer}. Also it has been recently 
shown by Pietro 
and Demaret\cite{Rpietro} that for constant scalar field equation of 
state, which is a good approximation for a tracker field solutions, 
the field equations and the conservation equations strongly constrain 
the scalar field potential. Most of the widely used potential for 
quintessence, such as inverse power law one, exponential or the cosine 
form, are incompatible with these constraints.

The CDM is in general considered to be a perfect fluid. However,
in some scenarios, certain physical processes can make
the CDM fluid effectively a dissipative one. In such a situation
the fluid has an effective pressure that is negetive.
Recently it has been proposed that the CDM must self interact in 
order to explain the detailed structure of the galactic halos 
\cite{CDM}. This self interaction will create a 
viscous pressure whose magnitude will depend on the mean free 
path of the CDM particles. In a recent work Chimento et.al 
have shown that a mixture of minimally coupled self interacting 
scalar field and a perfect fluid is 
unable to drive the accelerated 
expansion as well as solve the cosmic coincidence problem at the same 
time \cite{chimento}.
However, a mixture of a dissipative CDM with bulk viscosity and a 
minimally coupled self interacting scalar field 
can successfully achieve both features 
simultaneously. Also, as demonstrated in a recent paper by 
Zimdahl \etal ~\cite{zim} one can also have a negative $\pi$ if 
there exists an interaction which does not conserve particle numbers. 
This may be due to the particle production out of 
gravitational field. In this case, the CDM is not a conventional 
dissipative fluid, but a perfect fluid with varying particle number. 
Substantial particle production is an event that occurs in the 
early universe. But Zimdahl \etal have shown that even extremely small 
particle production rate can also cause the sufficiently negative 
$\pi$ to violate the strong energy condition.

In this paper we have used a minimally coupled scalar 
field with a self interacting potential together with a matter fluid 
having a dissipative pressure over and above its positive equilibrium  
pressure. We have not assumed any particular model for this negative 
pressure. Instead, we have investigated what 
kind effects it has in the expansion of the universe.
Unlike other works in scalar field cosmology with a dissipative 
pressure ~\cite{dis} we have neither assumed the behaviour of the 
scale factor nor have we assumed any specific form of the potential. 
Rather we have expressed all the variables in terms of what we call 
the `{\it generating function}'. For this we have followed the method 
described by Chimento et.al. \cite{chimento1} with some additional 
assumptions. 
We have proceeded with 
a particular choice of the generating function for which the potential 
is constructed using a combination of different power-law functions 
of $\phi$.
From the behaviour of decelerating parameter it has been shown that 
one can indeed generate both inflationary era in the early time and 
also an late time accelerating phase with a decelerating period in 
between. We have also investigated the stability and attractor 
structure of the general solutions of the field equations with this 
kind of potential and have found that for certain 
choices of the constants the solutions indeed exhibit attractor 
behaviour in the late times.


\section{Field Equations}
Let us consider a homogeneous, isotropic, spatially flat FRW universe
with a line element
\beqn
ds^2 = - dt^2 + a(t)^2~\sum_{i=1}^{3} dx_{i}^2,
\eeqn
where $a(t)$ is the scale factor. The energy density consists of
a massive scalar field with a self-interacting potential $V(\phi)$
\beqn
S_{\mu\nu} = \phi_{,\mu} \phi_{,\nu} - g_{\mu\nu}\left(\frac{1}{2}
\phi_{,\alpha} \phi^{,\alpha} + V(\phi)\right),
\eeqn
together with a dissipative fluid having bulk viscosity as the only 
dissipative term:
\beqn
T_{\mu\nu} = \rho u_{\mu} u_{\nu} + h_{\mu\nu} \left( p + \pi \right),
\eeqn
where $\rho$ is the energy density, $p$ is the equilibrium pressure, 
$\pi$ is the bulk viscous pressure, and 
$h_{\mu\nu} = g_{\mu\nu} + u_{\mu} u_{\nu}$
is the projection tensor.
\par
The independent field equations describing the system are 
\beqa
\label{kgeqn}
\ddot \phi + 3 H \dot\phi + \frac{dV}{d\phi} = 0,\\
\label{friedeqn}
3 H^2 = \frac{1}{2} \dot\phi^2 + V + \rho,\\
\label{conseq}
\dot\rho + 3 H \left( \rho + p + \pi \right) = 0
\eeqa
After some
straightforward calculations one can construct the equation
\beqn
\label{friedkgeqn}
2 \dot H = - \dot\phi^2 - \left( \rho + p + \pi\right).
\eeqn

For $\dot H \neq 0$, the system consists of three independent equations, viz., 
(\ref{conseq}),~(\ref{friedeqn}) and~(\ref{friedkgeqn}) and we have
six unknowns, $H, \phi, V(\phi), \rho, p$ and $\pi$. Hence, three 
constraint equations are required in order to have a closed system of 
equations. We first assume that $p$,$\rho$ and $\pi$ are related to 
${\dot \phi}$ by,
\beqn
\label{assumeqn1}
\rho + p + \pi = m \dot\phi^2,
\eeqn
where $m$ is an arbitrary constant. 
We emphasize that this
choice is considered purely because it makes the system of
equations simple to solve. We use this to see whether such a 
choice leads to any physically acceptable solutions. 
\par
The nature of the potential determines the evolution of $\phi$.
However, instead of choosing a particular form for $V(\phi)$, we
can alternatively describe the evolution of the scalar field by expressing
${\dot \phi}$ as a function of $\phi$ as,
\begin{equation}
\label{chosphi}
{\dot \phi}=F(\phi)
\end{equation}

where $F(\phi)$ is called the generating function. In this approach, we 
start with a specific  
trajectory, ${\dot \phi}=F(\phi)$, 
in phase space for $\phi$ and determine the potential that evolves
the scalar field in that manner. However, choosing a potential $V(\phi)$
or choosing a generating function, $F(\phi)$ is not completely equivalent. 
Choosing a form for $F({\phi})$
is more restrictive than choosing a potential due to the following reason.
A specific $F(\phi)$
refers to a specific class of initial conditions. Hence the set of
solutions 
represented by a specific form for $F(\phi)$ is a subset of the full set
of
solutions for the corresponding potential. Having constructed the
potential function, we study the evolution of the system in phase
space for a more general initial condition, {\it i.e.} ones which are
not restricted by any specific form for $F(\phi)$.

Using equation (\ref{assumeqn1}), equation (\ref{friedkgeqn}) and equation (\ref{chosphi})  one can write, 
\beqn
2\frac{dH}{d\phi} = - (m+1) F(\phi).
\eeqn
In this paper,
we first choose $F(\phi)$ and then
calculate the $V(\phi)$ that is consistent with our choice of $F(\phi)$ 
as well as with
the Einstein's equations. The solutions for which the phase space
trajectory 
is given by $\dot \phi=F(\phi)$ is a subset of the of the general class of
solutions for this potential.
(Later in subsection \ref{stability} 
we will draw the phase space trajectory for a set of general solutions
({\it i.e.} with different initial conditions) including the one for which 
$\dot \phi=F(\phi)$) and check the conditions for this to be a stable
solution.)
\par
For any given $F(\phi)$ which we term as the ``generating 
function'', one can, in principle, solve the system as follows:
\beqa
H(\phi) & = & -\frac{(m+1)}{2}\int F(\phi) d\phi~+~H_0, \label{Heqn}\\
\rho(\phi) & = & - 3 m \int H(\phi) F(\phi) d\phi +\rho_{0}, \label{rhoeqn}\\
a(\phi) & = & a_0~{\rm exp}\left[\int\frac{H(\phi)}{F(\phi)}d\phi \right] 
\label{aeqn}\\
V(\phi) & = & 3 H^2(\phi) - \frac{1}{2} F(\phi)^2 - \rho(\phi), 
\label{poteqn}\\
p + \pi & = & m F(\phi)^2 - \rho(\phi). \label{ppieqn}
\eeqa

where $H_{0}, \rho_{0}$ and $a_{0}$ are integration constant.With the assumption of an equation of state $p = p(\rho)$, one can also
calculate $\pi$. Hence, our main aim is to choose $F(\phi)$ 
properly to have some physically acceptable behaviour for different
variables. 

\section{Solutions for $F(\phi)=\omega \phi$}
\subsection{Exact solutions} \label{exact}

With this choice of the generating function, the differential equation
governing the time evolution of $\phi$ is
\begin{equation}
{\dot \phi}=\omega \phi. \label{phidiffeq}
\end{equation}
In this case, the exact solutions turn out to be,
\beqa
\phi(t) & = & \phi_0~{\rm exp}(\omega t),\label{phires}\\
H(t) & = & -\frac{(m+1)}{4} \omega \phi_0^2~{\rm exp}(2\omega t) +
H_0\label{Hres}\\
a(t) & = & a_0~{\rm exp}\left[-\frac{\phi_0^2 (m+1)}{8}~{\rm exp}(2\omega
t) + H_o t\right],
\label{ares} \\
q(t) & = & \frac{(m+1)}{2}\left[ \frac{\omega^2 \phi_0^2~{\rm exp}(2\omega
t)}
{\left( H_0 - \frac{(m+1)\phi_0^2 \omega~{\rm exp}(2\omega
t)}{4}\right)^2}\right] - 1,
\label{qres} \\
\rho(t) & = & \frac{3 (m+1) m \omega^2 \phi_0^4~{\rm exp}(4\omega t)}{16}
\nonumber \\
& - & \frac{3 m \omega H_0 \phi_0^2~{\rm exp}(2\omega t)}{2}+\rho_{0},
\label{rhores} \\
V(\phi) & = & (3 H_0^2-\rho_{0}) + \frac{3 (m+1) \omega^2 \phi^4}{16} -
\frac{\omega\phi^2}{2}\left( \omega + 3 H_0\right).\label{Vres}
\eeqa

For an expanding Universe we must have $H>0$. 
To ensure this, either 
$m<-1$ or $\omega<0$. We choose $\omega<0$. It may be noted
from equation (19) that the proper volume becomes zero at 
$t \longrightarrow -\infty$ and hence this is taken to be the 
initial time for our model.
The potential $V(\phi)$ given in equation (22) is the standard 
renormalizable tree level potential arising in the perturabative 
regime of quantum field theory.
With $\omega<0$, the shape of the potential depends on the factor 
($\omega+3H_0$). When 
$H_0>-\omega /3$, it is the most simplified version of the 
potential for the hybrid inflation \cite{lyth}. Similarly when 
$H_0<-\omega /3$, it is the inverted quadratic potential and 
has been discussed by many authors for inflationary models 
(see \cite{lyth} and references therein).
As has been pointed out before,
the solution represented by equation (\ref{phires}) 
is only one of the classes solutions (corresponding to the potential 
given in equation (\ref{Vres}))for which the the initial values of
$\dot{\phi}$ and $\phi$ are related by 
${\dot{\phi}}_{initial}=\omega{\phi}_{initial}$.
\vskip 0.1cm
Expressing the equation of state as, $p = (\gamma - 1)\rho$, 
where $1\leq\gamma\leq 2$, 
the bulk viscous pressure is given by, 
\beqa
\pi & = & m\omega\phi_0^2~{\rm exp}(2\omega t)\left(\omega + \frac{3\gamma
H_0}{2}\right)\nonumber\\
& - & \frac{3\gamma m(m+1)}{16}\omega^2\phi_0^4~{\rm exp}(4\omega t) -\gamma\rho_{0}.
\label{visc}
\eeqa
\subsection{Behaviour of the solutions}\label{behaviour}

\begin{figure}[ht]
\centering
\leavevmode\epsfysize=5cm \epsfbox{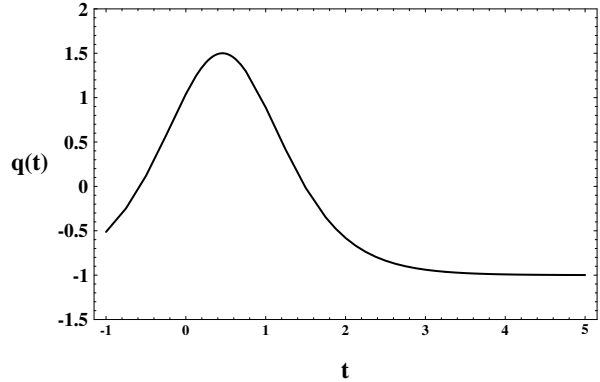}
\caption{Behaviour of the deceleration parameter  with time}
\label{figure1}
\end{figure}

\begin{figure}[ht]
\centering
\leavevmode\epsfysize=5cm \epsfbox{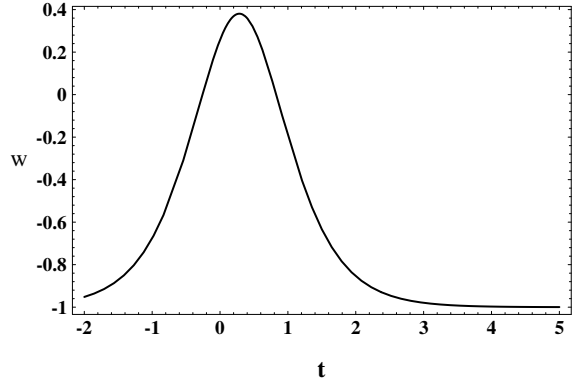}
\caption{Behaviour of ${\rm w}$ with time}
\label{figure 2}
\end{figure}

The behaviour of the  deceleration parameter is
shown in figure \ref{figure1}. 
The evolution of $q(t)$ has the basic feature that we require for an
acceptable form for the deceleration parameter.
$q(t)$ is negative to begin with resulting in an inflationary phase.
It increases and subsequently $q(t)$ becomes positive. At a later time,
it drops below $0$ and finally saturates at a constant value below zero.
This last feature of $q(t)$ at late times 
when it becomes a negative constant produces the present day accelerated 
expansion. In fact it asymptotically becomes a de Sitter Universe.
The behaviour of the equation of state 
for the scalar field  ${\rm w} = (\frac{1}{2}\dot\phi^2 -
V(\phi))/(\frac{1}{2}\dot\phi^2 + V(\phi))$ in figure (\ref{figure 2}) shows that
at late times ${\rm w} = -1$ which essentially 
supports the existence of a cosmological constant in late time. 

\begin{figure}[ht]
\centering
\leavevmode\epsfysize=5cm \epsfbox{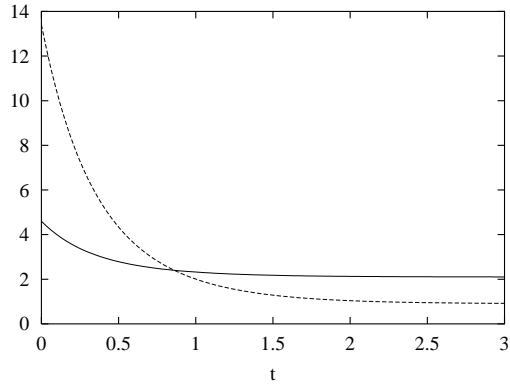}
\caption{Behaviour of $\rho$ (dashed)and $\rho_{\phi}$ (solid)  with time}
\label{figure4}
\end{figure}

An acceptable model should also satisfy constraints arising from
cosmological nucleosynthesis and from structure formation.
In order to keep these intact, the matter energy density should dominate over
the energy density of the scalar field in the early universe. 
However, at late times the 
scalar field energy density should be more than that due to matter 
so that one can explain the {\it missing energy} of the universe.
In figure (\ref{figure4}) we have plotted the energy density for the matter 
$(\rho)$ and for the scalar field $(\rho_{\phi})$. 
We see the above constraints are satisfied.
The matter energy density is dominant in the early time
and so nucleosynthesis and structure formation are unaffected.
But at late time  $\rho_{\phi}$ decreases more slowly than 
$\rho$ and hence ultimately it becomes greater than $\rho$ 
This explains the missing energy associated with accelerated expansion 
of the universe.  It can also be seen that although in early period the energy densities are of different order of magnitude but in late time they are of the same order. This gives a dynamical solution to the cosmic coincidence 
problem.
Also at late times the ratio of these 
two energy densities becomes constant showing  the tracking behaviour. 
We have also plotted the two density parameter $\Omega_{m}$ and 
$\Omega_{\phi}$ in figure 4.

\begin{figure}[ht]
\centering
\leavevmode\epsfysize=5cm \epsfbox{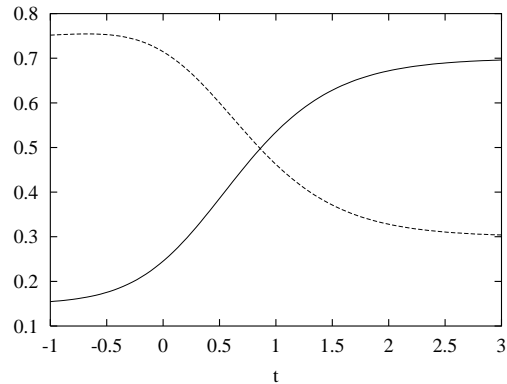}
\caption{Behaviour of $\Omega_{m}$ (dashed)  and $\Omega_{\phi}$ (solid)  with time}
\label{figure5}
\end{figure}
\begin{figure}[ht]
\centering
\leavevmode\epsfysize=5cm \epsfbox{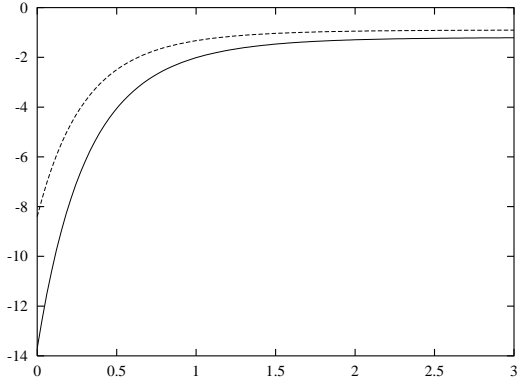}
\caption{Behaviour of viscous pressure for $\gamma=1$~ (dashed)and for 
$\gamma=4/3$ (solid).
with time}
\label{figure6}
\end{figure}

The variation of the viscous pressure $\pi$  is shown in \ref{figure6}. 

\subsection{Stability and attractor structure}\label{stability}
The solutions we discussed were those for which $F(\phi)=\omega \phi$. In
other
words, for these set of solutions the initial conditions are constrained
by the relation
$\dot{\phi}_i=\omega \phi_i$, where $\phi_i$ and ${\dot \phi}_i$ 
are the initial values of $\phi$ and ${\dot \phi}$, respectively.
These solutions are significant only if these are stable, {\it i.e.},
only if more general initial conditions asymptotically evolve towards these
phase space trajectories.

This solution corresponds to the case
when the expression for the Hubble parameter ($H(t)$), scale
factor ($a(t)$), deceleration parameter ($q(t)$) and the density
($\rho(t)$)
given by equations (\ref{Hres}), (\ref{ares}), (\ref{qres}) and
(\ref{rhores}).
In this section we proceed to investigate whether or not the solutions
described are stable and whether there are any constraints on the parameters
for this.

The general equation of motion for the potential given in equation 
(\ref{Vres}) is 
\beq
{\ddot \phi}=-3H{\dot \phi}
-\frac{3(m+1)}{4}{\omega}^2{\phi}^3+{\omega}^2\phi+3H_0\omega \phi
\eeq
Together with the equation for the evolution of $H(t)$, we can 
write the following set of three coupled differential equations,
\beqa
{\dot \phi}&=&x\\
{\dot
x}&=&-3Hx-\frac{3(m+1)}{4}{\omega}^2{\phi}^3+{\omega}^2\phi+3H_0\omega
\phi \\
{\dot H}&=&-\frac{(m+1)}{2}x^2\\
\eeqa

By equating
the right hand sides of the equations to $0$ we obtain the following three
critical points in the ($\phi, {\dot \phi}$) plane.
\beqa
(0,0), \left(\sqrt{\frac{4(\omega + 3H_0)}{3(m+1)\omega}},0\right), 
\left(-\sqrt{\frac{4(\omega + 3H_0)}{3(m+1)\omega}},0\right)\nonumber.
\eeqa
The position of the critical points in the phase diagram changes with the
value of $H_0$.
Further, at $\omega=-3H_0$ all the three critical points merge.
There is a transition in the nature of stability of the trajectories in 
$(\phi,{\dot \phi})$
space at this value of $H_0$. We discuss below the nature of the
trajectories for different values of 
$H_0$.
First of all, we should note that $\omega<0$. 

\begin{figure}[ht]
\centering
\leavevmode\epsfysize=7cm \epsfbox{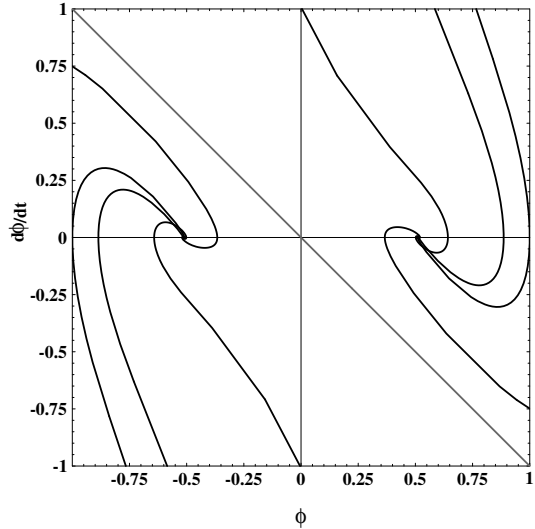}
\caption{Phase space trajectories for $H_0 = 0.2$}
\label{figure7}
\end{figure}

\begin{figure}[ht]
\centering
\leavevmode\epsfysize=7cm \epsfbox{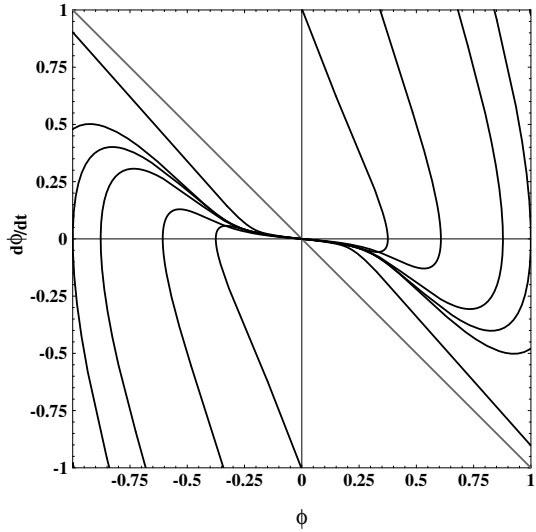}
\caption{Phase space trajectories for $H_0 = 0.4$}
\label{figure8}
\end{figure}

In figures (\ref{figure7}) and (\ref{figure8}), we have plotted the 
phase space trajectories for the scalar field
with a variety of initial conditions that are not constrained by the 
particular choice of the generating function.
Figure \ref{figure7} corresponds to the case of $H_0=0.2$. 
As we have chosen $\omega$ to be $-1$, this is a case in which the 
potential has a 
local maximum. On the other hand, figure \ref{figure8} 
corresponds to the case of $H_0=0.4$ in which
$\phi=0$ is a minimum for the potential.
In the first case the phase space trajectories converge to either of the
two
non-zero values of $\phi$ and is not an attractor like solution as the asymptotic nature of the solution depends on the initial conditions. But in the second case, however, the trajectories 
converge to $\phi=0$, ${\dot \phi}=0$ independent of the initial conditions  and ,the de-Sitter solution  is an attractor for this
case.
\section{Conclusion}
The most important conclusion is that models with a self-interacting scalar 
field and a matter fluid having a negative pressure in addition to its 
positive equilibrium pressure can produce a scenario for the cosmological 
evolution in which one can have an inflationary phase to begin with, an 
accelerated phase at late times (like the present era) and a decelerating 
phase in-between. 
Recently Lopez and Matos~\cite{matos} have shown that this kind of complete 
history for the scale factor can be described by a hyperbolic potential. 
But the physical origin of such potential is still not well known. 
But here we have shown that such kind of behaviour for the scale 
factor can be generated with a potential given in (22),which has been widely used by many authors for inflationary models.
\par
The behaviour of $\rho$ and $\rho_{\phi}$ in 
our model shows that although in early universe, $\rho$ 
is greater than $\rho_{\phi}$ which is necessary for different physical 
phenomena like nucleosynthesis and structure formation etc, in the late 
times, $\rho_{\phi}$ starts dominating. This feature explains 
the missing energy 
density and also the ration of two energy densities becomes a constant 
in late time showing the ``{\it tracking nature}''.

One should note that both the assumptions (8) and (16) play crucial role in our model. Given a barotropic equation of state between $\rho$ and $p$ one can not assume (8) and (16) at the same time if the dissipative pressure $\pi$ is zero in our model as t
hat will lead to an over-determined problem ( the number of unknowns will be less than the number of independent equations). Even if one assumes these two condition one can check that will lead to an negative equilibrium pressure which is not desirable. H
ence the existence of dissipative pressure also plays an important role in our model.

We have also studied the general equation of motion for the scalar field 
(equation (24)) for the potential (22) and have shown that for the choices 
of constant for which the potential is minimum at $\phi=0$, the phase 
space diagram exhibit a attractor behaviour towards the asymptotic de-sitter 
solution.

We want to mention that previously tracker and attractor solutions have 
been studied for scalar fields having inverse power law, exponential, 
cosine potential. But in all of these cases the equation of state $w$ 
is a constant in radiation era as well as in a matter dominated era. 
It was later shown by Pietro and Demaret ~\cite{Rpietro} that these 
kind of potential with a constant $w$ is not consistent with the field 
equations. In our case, the equation of state for the scalar field $w$ 
is not a constant but it varies with the cosmic evolution and approaches 
towards -1 asymptotically showing the existence of a cosmological constant 
in late times.


\end{document}